\documentclass[a4paper,aps,prl,twocolumn,groupedaddress]{revtex4}
\usepackage{graphicx}
\usepackage{dcolumn}% Align table columns on decimal point
\usepackage{bm}% bold math

\begin{document}
\title{Suppression of spin relaxation in an InAs nanowire double quantum dot}
\author{A.~Pfund, I.~Shorubalko, K.~Ensslin and R.~Leturcq}

\address{Solid State Physics Laboratory, ETH Z\"urich, 8093 Z\"urich, Switzerland\\
E-mail: leturcq@phys.ethz.ch }

\begin{abstract}
We investigate the triplet-singlet relaxation in a double quantum dot defined by top-gates in an InAs nanowire. In the Pauli spin blockade regime, the leakage current can be mainly attributed to spin relaxation. While at weak and strong inter-dot coupling relaxation is dominated by two individual mechanisms, the relaxation is strongly reduced at intermediate coupling and finite magnetic field. In addition we observe a charateristic bistability of the spin-non conserving current as a function of magnetic field.
We propose a model where these features are explained by the polarization of nuclear spins enabled by the interplay between hyperfine and spin-orbit mediated relaxation.
\end{abstract}

\maketitle
Coherent spin manipulation \cite{PettaScience2005,Koppens:2006fk} has been demonstrated on quantum dots in GaAs heterostructures, benefiting from the high quality of these systems and the highly developed manufacturing technology.  Alternative materials such as InAs may offer advantageous spin properties. The large electronic g*-factor could allow for easier spin control with magnetic fields, while the enhanced spin-orbit interactions motivated recent proposals for purely electrical manipulation of individual spins \cite{golovach-2006,flindt-2006}.

Various relaxation processes limit the lifetime of spin states. In GaAs quantum dots, the dominating mechanisms are the spin-orbit interaction \cite{golovach:016601,PhysRevB.61.12639,PhysRevB.64.125316,stano:186602,destefani:115326} and the hyperfine interaction with nuclear spins \cite{PhysRevB.65.205309,PhysRevLett.88.186802,coish:125337,Jouravlev:2006kx,PhysRevB.64.195306,PhysRevB.66.155327}. The question of spin relaxation is particularly relevant in InAs, where the strong spin-orbit interaction and the large nuclear spin of In are expected to lead to a strong spin relaxation. 

As shown in \cite{Koppens:2005qy,johnsonPulse}, spin relaxation can be probed by the leakage current through a double quantum dot (DQD) in the Pauli spin blockade regime \cite{Ono_rectification}. We consider three different leakage processes: (i) cotunneling through the DQD (ii) relaxation due to spin-orbit interaction and (iii) relaxation as a result of hyperfine interaction with the nuclei of the host material. 
When studying the kinetics of different processes occuring in parallel, it is common to consider a global reaction rate as the sum of the individual rates. Here we show that the interplay between the different spin relaxation mechanisms can lead to more complex phenomena, and even to a pronounced reduction of the global relaxation rate.

We investigate the singlet-triplet relaxation as a function of magnetic field in a DQD fabricated in an InAs nanowire (NW). At both weak and strong coupling between the two dots, strong relaxation is found. At intermediate coupling, we observe suppressed relaxation accompanied by a bistable behavior in magnetic field and gate detuning. We suggest a model where this suppression is explained by a dynamic nuclear spin polarization and supported only in the presence of different independent relaxation paths.
\begin{figure}
\includegraphics[width=0.45\textwidth]{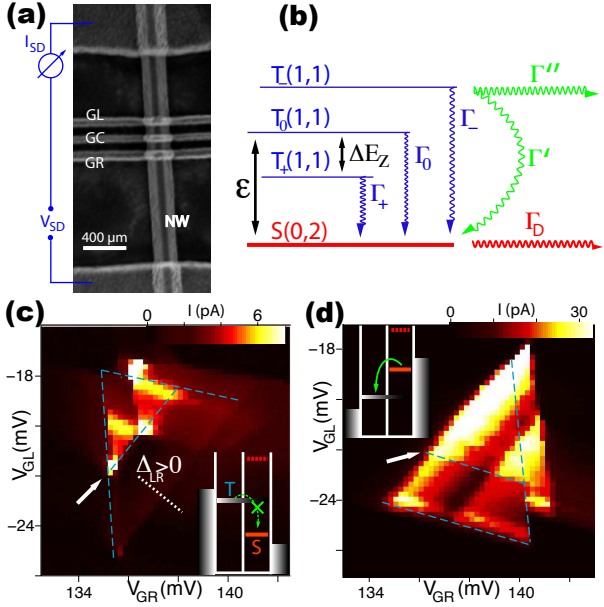}
\caption{a) Scanning electron microscope image of a device equivalent to the one measured. The current $I_{SD}$ for an applied source-drain voltage $V_{SD}$ is measured between the outer contacts to the nanowire (NW). Voltages on the top-gates GL,GR,GC create a double quantum dot. b)  Level scheme and relaxation rates inside the spin blockaded region. A small magnetic field splits the triplets by $\Delta E_Z$. Fluctuating nuclear spins lead to relaxation of the triplets to the singlets with rates $\Gamma_{0,\pm}$ over the splitting $\epsilon$. The singlet $S(0,2)$ is coupled to the drain lead with the rate $\Gamma_D$. Alternative processes such as cotunneling with a rate $\Gamma^{\prime\prime}$ and relaxation mediated by spin orbit interaction ($\Gamma^{\prime}$) are sketched for $T_-(1,1)$. c) Current $I_{SD}$ as a function of gate voltages $V_{GL}$ and $V_{GR}$ for $V_{SD}=-3$ mV. Spin blockade suppresses current in the base region of the triangle. Variation of $V_{GL},V_{GR}$ along the dotted white line detunes the levels in the dot with respect to each other by an energy $\Delta_{LR}$. At large detuning, current can flow via $T(0,2)$ (red dashed line in inset, white arrow in main panel). d) For reverse bias $V_{SD}=+3$\,mV, current can always flow via \mbox{$(0,1)\rightarrow (0,2) \rightarrow (1,1) \rightarrow (0,1)$} (inset). The white arrow indicates the onset of transport via $T(0,2)$.}
\end{figure}

The InAs NWs are catalytically grown using Metal Organic Vapor Phase Epitaxy \cite{Seifert:2004lr}, and are subsequently transferred to an oxidized silicon substrate. Ohmic contacts and top-gates are then created in a two-step lithographic process \cite{pfund:252106}, yielding devices as shown in Fig.\,1(a). We perform transport measurements in a dilution refrigerator at a base temperature of $30$\,mK unless otherwise stated. A magnetic field is applied perpendicular to the wire axis.
Two quantum dots in series are formed along the NW by applying voltages to the top-gates GL, GC and GR. In the Coulomb blockade regime, the number of electrons $(n,m)$ in the left/right dot can be tuned with the gates GL/GR, while the gate GC is used to adjust the coupling between the dots. For finite bias voltage $V_{SD}$, sequential electron transport via the dot occupation cycle \mbox{$(n,m)\rightarrow (n+1,m) \rightarrow (n,m+1) \rightarrow (n,m)$} is energetically allowed only for gate voltages in a triangular region of the $V_{GL}$-$V_{GR}$-plane \cite{vanderWiel01,pfund:252106}. This is demonstrated in Figs.\,1(c) and (d) where the current is measured for two opposite bias voltages. Adjusting the voltages $V_{GL}$ and $V_{GR}$ we can either tune the global energy in both dots (along the base of the triangles), or detune the levels in the dots with respect to each other by an energy $\Delta_{LR}$ (along the doted line in Fig.\,1(c)).

While for positive $V_{SD}$ (Fig.\,1(d)) a large current is measured in the whole triangle, the current is suppressed close to the base of the triangle for negative bias (Fig.\,1(c)). Similar results have been observed in GaAs DQDs containing two electrons \cite{Ono_rectification}, and attributed to spin blockade (SB) due to the Pauli exclusion principle. The two electrons can form either singlet or triplet states, for both occupations $(1,1)$ and $(0,2)$. Because tunneling preserves spin, the evolution from a $(1,1)$ triplet to the $(0,2)$ singlet is forbidden, which suppresses the current (inset Fig.\,1(c)). If the detuning $\Delta_{LR}$ exceeds the singlet-triplet splitting ($S(0,2)$-$T(0,2)$) of the drain-side dot, electrons can escape also from the $(1,1)$-triplets via $T(0,2)$ and a finite current is observed (white arrow in Fig.\,1(c)). For reverse bias, electrons can always pass from the $(0,2)$ to the $(1,1)$ states (inset Fig.\,1(d)). The splitting between $S(0,2)$ and $T(0,2)$ is consistently seen as a step in current.

In our device, we are not able to decrease the number of electrons down to two. However, SB was observed in GaAs DQDs also for charge transitions with higher even numbers of electrons \cite{johnson:165308}. In InAs, interaction effects are weaker because of the small effective mass ($m^*\approx 0.02m_0$). Therefore spin pairing is more likely to occur and the spin-less core of electrons can be ignored. To further check the consistency with SB, we have measured the same effect when changing the number of electrons in the DQD by two, while SB is not observed when changing this number by one.

To investigate the relaxation processes, we measure the leakage current through the SB as a function of magnetic field and level detuning $\Delta_{LR}$. The observed relaxation changes drastically for different coupling between the two dots. We can tune the tunnel coupling energy over a wide range in our device \cite{pfund:252106} and estimate values of $0.15$-$0.25$\, meV for the regimes considered here.

For weakly coupled dots, we observe a significant leakage through the SB at low fields, which is reduced for higher fields (not shown). Our data are in good agreement with results on weakly coupled lateral DQDs in GaAs \cite{Koppens:2005qy}, where the relaxation was explained by the environment of fluctuating nuclear spins.
\begin{figure}
\includegraphics[width=0.45\textwidth]{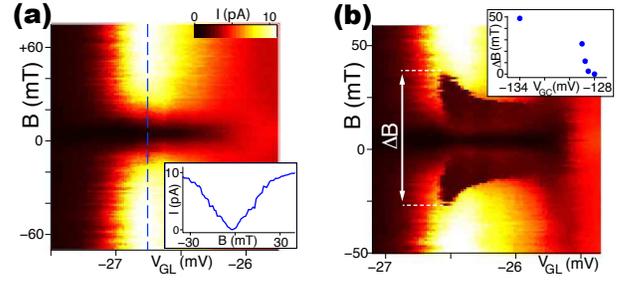}
\caption{(a) Color plot of $I_{SD}$ for $V_{GC}=-128$\,mV. The gate voltages are detuned along the white dotted line as shown in Fig.\,1(b) and the B field is stepped. Inset: magnetic field dependence of the current along the blue dashed line in the main panel. 
(b) Same measurement as in (a), but with slightly deceased inter-dot coupling ($V_{GC}=-129$\,mV). Inset: size of the low leakage region in units of the magnetic field for different coupling, tuned by $V_{GC}$.}
\end{figure}

The situation is reversed for strong inter-dot coupling. In Fig.\,2(a), the leakage current is plotted as a function of detuning and magnetic field. The gate voltages are varied along the dotted detuning line in Fig.\,1(c). Spin blockade is observed around \mbox{$B=0$\,mT} and lifted for a small finite field. The inset shows the field dependence of the leakage current along the dashed line in the main panel. After a monotonic increase for small fields, the leakage current saturates to a field independent value probably limited by the coupling to the leads.

Tuning to intermediate coupling, we observe a complex behavior. In Fig.\,2(b), the measurement of Fig.\,2(a) is repeated with slightly decreased coupling. The leakage current is suppressed by an order of magnitude in a sharply defined region of magnetic field and detuning. From the magnitude of the current ($I\sim 1$\,pA), we estimate a relaxation time $T_1^* > e/I\sim 100$\,ns. The suppression only exists up to a critical coupling strength ($V_{GC}=-128$\,mV). Below this value, the dimension of the low-current region varies strongly with $V_{GC}$ (Fig.\,2(b) inset). 

The reduced leakage current exhibits a pronounced bistability. Fig.\,3 shows measurements corresponding to Fig.\,2, but sweeping the magnetic field in one direction and stepping the detuning after each sweep. The region of reduced relaxation is drastically extended in the sweep direction of the field. We find no dependence on the sweep rate within the experimental values (0.5 mT/min -- 300 mT/min). After accessing a point in the extended region (labeled X in Fig.\,3(b)), the low-current state is stable for hours. The high leakage current is switched on again, if the gate voltages are tuned outside the region of suppression and back along the dashed line (Fig.\,3(b) inset).

In Fig.\,4(c), current traces for up (green) and down (black) sweeps of the magnetic field along the dashed line in Fig.\,3(a) are shown. The bistable suppression abruptly vanishes above a critical temperature of $T_{cr}\approx 600\,$mK, see Fig.\,4(d). At high temperatures, the current traces are similar to the ones for high coupling taken along the dashed line in Fig.\,2(a) (dotted lines in Fig.\,4(c)).

We now discuss our data in view of the relaxation mechanisms sketched in Fig.\,1(b). (i) Cotunneling gives rise to a leakage background as soon as a level in one of the dots is within the bias window. This process depends on the coupling to the leads, and we measure almost no dependence on magnetic field in the range considered here (up to several hundred mT).
(ii) In the presence of spin-orbit interactions, no fixed spin quantization axis exists and singlets and triplets are mixed \cite{PhysRevB.61.12639,PhysRevB.64.125316}. Spin relaxation can then occur via electron-phonon interactions. This mechanism should not lead to relaxation at zero magnetic field due to time reversal symmetry and the rate is expected to increase with a power-law for finite field \cite{golovach:016601,stano:186602,Hanson:2005uq,meunier-2006}
(iii) Finally, the hyperfine interaction depends on the overlap of the electronic wavefunction with the nuclei of the host material. In a DQD, electrons in the separated dots experience uncorrelated local fields, which leads to a mixing of the $(1,1)$ triplets with the singlet \cite{PhysRevB.65.205309,PhysRevLett.88.186802,coish:125337}. For weakly coupled dots, the resulting triplet-singlet relaxation is strong at small magnetic fields, while it is suppressed for higher fields \cite{Jouravlev:2006kx}. This magnetic field dependence is in contrast to the relaxation due to spin-orbit interaction.

At strong coupling (Fig.\,2(a)), the suppression of the leakage current at low magnetic field and its continuous increase with $B$ is very different from what has been observed in GaAs DQDs for weak, intermediate and strong coupling \cite{Koppens:2005qy,ono:256803}. However, the behavior is consistent with the magnetic field dependence of spin relaxation due to spin-orbit coupling. Strong spin-orbit interaction is indeed expected in InAs. Furthermore, we observe a pronounced anticrossing of the $S(0,2)$ and $T(0,2)$ states at $B\approx2.5$\,T (not shown), which has been studied in a similar system and can be explained by spin-orbit coupling \cite{bulaev:205324,FuhrerComm}. We thus attribute the leakage current at strong coupling mainly to relaxation by spin-orbit interaction.
\begin{figure}
\includegraphics[width=0.45\textwidth]{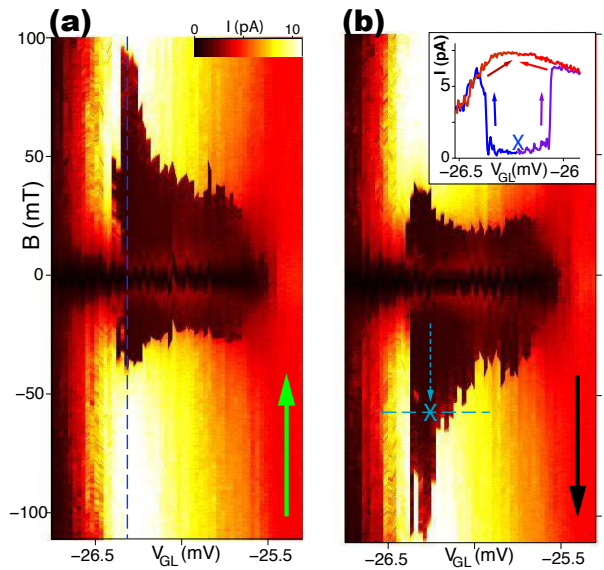}
\caption{Measurements corresponding to Fig.\,2(b), but sweeping the magnetic field and stepping the gate voltages after each trace. (a) Up-sweeps $-120$\,mT to $+100$\,mT (b) Down-sweeps. Inset: Current traces for variation of the gate voltages starting from point X demonstrating the bistability. The high-current state is stable, once the gates are tuned out of the region of suppressed relaxation.}
\end{figure}

At intermediate coupling, the observed bistability of the leakage current differs strongly from the continuous magnetic field dependences at weak and strong coupling. A similar bistability has been detected in the spin blockaded regime in GaAs DQDs \cite{ono:256803,Koppens:2005qy}, as well as by optical pumping in InGaAs dots \cite{maletinsky-2006,Braun:0607728}. In both cases, the effect was attributed to the dynamic polarization of nuclei through the hyperfine interaction. While the bistable behavior in our system is different from the one reported in GaAs DQDs \cite{ono:256803,Koppens:2005qy}, we believe that the strong reduction of leakage current is related to dynamic polarization of the nuclei.

To explain our results, we propose a model where a net nuclear polarization can be achieved only when hyperfine-induced processes occur in parallel with alternative relaxation paths, which we attribute to spin-orbit interaction. The polarized nuclei then create an effective magnetic field $B_N$ opposite to the external field $B$, which suppresses the relaxation at larger fields.

Recent theoretical work addresses the question how nuclear spin polarization can emerge in a spin blockaded DQD \cite{rudner-2006}. The transitions $T_{\pm} \rightarrow S$ involve a spin-flip for an electron. If mediated by the hyperfine interaction \cite{PhysRevB.64.195306,PhysRevB.66.155327}, this transfers a spin quantum from the electron to the nuclei (so-called flip-flop process). The flip-flop rates $\Gamma_{\pm}$ depend on the energy splitting between $T_{\pm}$ and the singlet, and are maximum for aligned energy levels. As shown in Fig.\,4(a), the splitting varies with the detuning $\Delta_{LR}$ between the dot levels and the magnetic field that splits $T_\pm$ from $T_0$ \cite{PettaScience2005,Koppens:2005qy}. The singlets $S(1,1)$ and $S(0,2)$ are hybridized by the tunnel coupling and form two branches.
Yet even if one of the rates is much higher than the others, no nuclear polarization is build up since electrons enter into the states $T_+$ or $T_-$ with the same probability. A net polarization is expected only if alternative paths from $T$ to $S$ (e.g. mediated by spin-orbit coupling) allow the fastest rate to dominate the net relaxation \cite{rudner-2006}.
\begin{figure}
\includegraphics[width=0.45\textwidth]{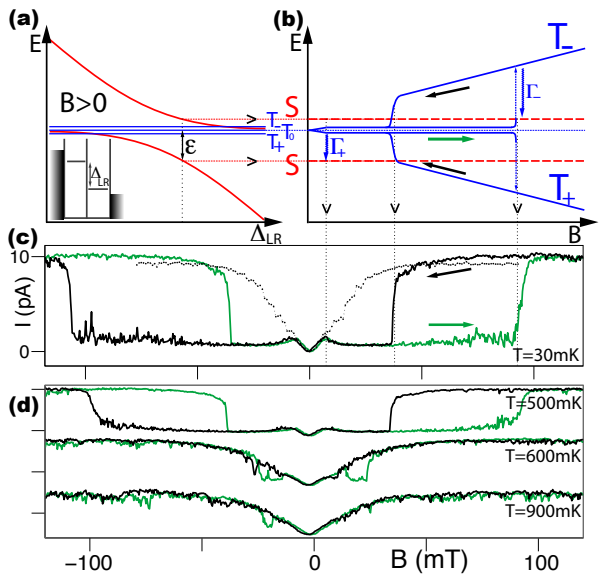}
\caption{(a) Dependence of the singlet branches S and triplets $T_{0,\pm}(1,1)$ on the detuning $\Delta_{LR}$ of the dot levels. (b) Evolution of the triplet energies at fixed $\Delta_{LR}$ for up- and down-sweeps of the magnetic field, including the influence of dynamic nuclear spin polarization. Arrows indicate the sweep direction. (c) Current trace along the dashed line in Fig.\,3(b). Up-sweeps (green) and down-sweeps (black). The dotted curves are taken at stronger coupling (Fig.\,2(a)). (d) Current along the dashed line in Fig.\,3(b) for different temperatures.}
\end{figure}

The scenario we propose is depicted in Fig.\,4(b). For small fields, $T_+$ approaches the lower singlet branch S, while $T_-$ is shifted away in energy. The relaxation $T_+ \rightarrow S$ thus dominates. During this process, an electron is flipped from parallel to antiparallel alignment with the external field. Correspondingly, an increasing amount of nuclear spins are flipped parallel to the field. Due to the negative g-factor in InAs, the nuclear field $B_N$ therefore counteracts the external field \cite{Hanson01}. Further Zeeman splitting is inhibited and the triplets evolve along the inner traces in Fig.\,4(b) (green arrow). The leakage current is thus sustained at the low field value $\sim 1$\,pA (Fig.\,4(c)). For further increasing field, a saturation for the nuclear polarization is reached and the triplets split more. As soon as $T_-$ crosses the upper singlet branch, the $T_- \rightarrow S$ relaxation becomes energetically possible. This flip-flop process builds up a nuclear field that supports the external field, which leads to a sharp rise of the Zeeman splitting and to a step in current. Decreasing the B field, the high current state is stable until $T_+$ crosses $S$ again and the nuclei reduce the effective field for the electrons.

The bistable behavior vanishes above a critical value for the coupling (Fig.\,2(b) inset). This can be understood when solving the rate equation for the flip-flop process in a simplified model with $\textrm{spin-}\frac{1}{2}$ nuclei, which leads to a condition for the creation of a nuclear spin polarization \cite{rudner-2006}:
$$\left(\epsilon+\mu_B g^* B_{N,t}\right)^2+(h\Gamma_D)^2 < (\mu_B g^* B_{N,t})^2.$$
Here, $\epsilon$ is the \mbox{$S$-$T_0$} splitting, $\Gamma_D$ is the coupling of $S(0,2)$ to the drain lead and $B_{N,t}$ the total nuclear field produced by the polarization of all nuclei. This condition shows that nuclear spin polarization is not possible for a too large splitting $\epsilon$. Increasing the coupling leads to an increase of the \mbox{$S(1,1)$-$S(0,2)$} anticrossing and thus to an increase of $\epsilon$.

The insensitivity on the magnetic field sweep rate suggests that nuclear spin relaxation times are smaller than $1$\,s, which is significantly shorter than in GaAs \cite{Koppens:2005qy,ono:256803}. Fast nuclear spin relaxation has also been observed in InGaAs QDs by optical pumping \cite{Braun:0607728}. A puzzle is the abrupt suppression of the bistable behavior above $T_{cr} \approx 600$ mK, which has not been observed in GaAs quantum dots \cite{ono:256803}. 

In conclusion we investigated spin relaxation by the leakage current through a DQD in the regime of spin blockade. We showed, that the interplay of two independent relaxation mechanisms leads to suppressed leakage compared to the individual contributions.

\begin{acknowledgments}
We thank J.~Elzerman, T.~Ihn, A.~Imamoglu, P.~Maletinski, J.~Taylor for usefull discussions, M.~Borgstr\"om, E.~Gini for advice in nanowire growth and F.~Gramm, E.~M\"uller for TEM imaging. We acknowledge financial support from the ETH Zurich.
\end{acknowledgments}
\bibliography{Pfund_SupprRelax}
\end{document}